\documentclass[a4paper,14pt]{article}
\usepackage{soul}
\usepackage[T2A]{fontenc}
\usepackage[english]{babel} 
\usepackage[linktocpage,unicode]{hyperref}
\usepackage{csquotes,graphicx,fancyhdr}
\usepackage{setspace,longtable,graphicx,hyphenat,hyperref,fancyhdr,ifthen,enumitem,amsmath}
\usepackage[dvipsnames]{xcolor}
\usepackage[export]{adjustbox}
\usepackage[linktocpage,unicode]{hyperref}
\hypersetup{unicode=true,colorlinks=true,linkcolor=black,citecolor=blue,urlcolor=blue}
\usepackage{url}
\usepackage{amssymb,amsmath}
\usepackage{geometry}
\usepackage[style=gost-numeric,bibencoding=auto,backend=biber,sorting=none,language=auto,autolang=other,doi=false,url=false]{biblatex}
\AtEveryBibitem{\clearfield{day}
	\clearfield{month}
	\clearfield{endday}
	\clearfield{endmonth}}
\bibliography{Ru-Article_6} 
\usepackage{multido}
\newcommand{\e}{\text{e}}
\newcommand{\n}{\text{n}}
\newcommand{\D}{\text{D}}
\newcommand{\M}{\text{M}}
\newcommand{\N}{\text{N}}
\newcommand{\K}{\text{K}}

\newcommand{\CC}{\text{C}}
\newcommand{\LL}{\text{L}}
\newcommand{\HH}{\text{H}}
\newcommand{\Pl}{\text{Pl}}
\newcommand{\cc}{\text{c}}
\newcommand{\eff}{\text{eff}}
\makeatletter
\ams@newcommand{\vardot}[2]{%
  {\mathop{#2\kern0pt}\limits^{\vbox to-1.4\ex@{\kern-\tw@\ex@
   \hbox{\normalfont\multido{}{#1}{.}}\vss}}}}
 \makeatletter
\def\@fnsymbol#1{\ensuremath{\ifcase#1\or \dagger\or  *\or \dagger\dagger
   \or \ddagger\ddagger \else\@ctrerr\fi}}
    \makeatother
\makeatother

\def\nq{\hspace*{-1em}}          

\def\nhq{\hspace*{-0.5em}}       

\def\cm{\hspace*{1cm}}          
\def\inch{\hspace*{1in}}

\def\qq{\qquad} 
\def\d{\partial}

\def\ME{\mbox{$\M_{\rm E}$}}
\def\MJ{\mbox{$\M_{\rm J}$}}
\def\const{{\rm const}}            
\def\eps{\varepsilon}

\def\then{\ \ \Longrightarrow\ \ } 
\def\beq{\begin{equation}}
\def\eeq{\end{equation}}
\def\lal{&& {}\nq}
\def\bear{\begin{eqnarray}}            
\def\bearr{\begin{eqnarray} \lal}
\def\ear{\end{eqnarray}}               
\def\earn{\nonumber \end{eqnarray}}
\def\nn{\nonumber\\ {}}                

\def\nnn{\nonumber\\ \lal }            

\def\yy{\\[5pt] {}}                    
\def\yyy{\\[5pt] \lal }
\def\eql{&\nhq = &\nhq}                    

\def\rf{\eqref}
\def\eq{Eq.\,}
\def\eqs{Eqs.\,}

\def\det{\mathop{\rm det}\nolimits}
\def\mn{_{\mu\nu}}
\def\MN{^{\mu\nu}}

\def\og{{\overline g}}
\def\oR{{\overline R}}

\def\ssph{static, spherically symmetric}

\def\email#1#2{\footnotetext[#1]{e-mail: #2}\addtocounter{footnote}{1}}
\usepackage{color}

\title{\bf Multi-scale hierarchy from multidimensional gravity}

\author{
Kirill A. Bronnikov$^{a,b,c,1}$, Arkady A. Popov$^{d,2}$, Sergey G. Rubin$^{c,d,3}$ 
        }

\date{\small\it
$^a$  Center fo Gravitation and Fundamental Metrology, VNIIMS,
		Ozyornaya ulitsa 46, Moscow 119361, Russia\\
$^b$  Institute of Gravitation and Cosmology, RUDN University, 
		ulitsa Miklukho-Maklaya 6, Moscow 117198, Russia \\
$^c$  National Research Nuclear University MEPhI (Moscow Engineering Physics Institute),\\ 
            Kashirskoe shosse 31, Moscow 115409, Russia \\
$^d$  N.I. Lobachevsky Institute of Mathematics and Mechanics,
	Kazan  Federal  University, \\
	Kremlyovskaya ulitsa 18,  Kazan 420008,  Russia}

\begin{document}
\maketitle
\email{1}{kb20@yandex.ru}
\email{2}{arkady\_popov@mail.ru}
\email{3}{sergeirubin@list.ru}

\begin{abstract}
We discuss the way of solving the hierarchy problem. We show that starting at the Planck scale, 
    the three energy scales --- inflationary, electroweak and the cosmological ones can be restored. The 
    formation of small parameters is proposed that leads to a successful solution of the problem. 
    The tools involved in the process are $f(R)$ gravity and inhomogeneous extra dimensions.
    Slow rolling of a space domain from the Planck scale down to the inflationary one gives rise 
    to three consequences: an infinite set of causally disconnected domains 
    (pocket universes) are nucleated; quantum fluctuations in each domain produce a variety of different fields and 
    an extra-dimensional metric distribution; these distributions are stabilized at a sufficiently
    low energy scale.  
\end{abstract}

\section{Introduction}

 Assuming that the Universe has been formed at the Planck scale,
 it is naturally implied that its initially formed parameters are of the order of 
 the same scale. The essence of the Hierarchy problem is the question: Why are 
 the observable low-energy physical parameters so small as compared to those of 
 the Planck scale? How did Nature manages to decrease the parameter values so substantially? 

 There are at least four important energy scales during evolution of the Universe: the Planck scale 
 ($\sim 10^{19}$\,GeV) at which our Universe cannot be described by classical laws; the 
 inflationary scale ($\sim 10^{13}$\,GeV) where our horizon has appeared, the electroweak scale  
 ($\sim 10^2$\,GeV), and the cosmological scale specified by the cosmological 
 constant ($\sim 10^{-123}$\,GeV$^4$) (CC).

 According to the inflationary paradigm, the physical laws are formed at high energies \cite{Brandenberger:2006vv,Tegmark:2005dy}, where the Lagrangian structure is yet unknown. 
 The physics has 
 been established at the energy scale $M$ higher than the inflationary one, $E_I \sim 10^{13}$ GeV, 
 see  \cite{Loeb:2006en,Ashoorioon:2013eia} in this context. We study the way of substantially 
 decreasing the physical parameters at the three scales mentioned above assuming natural values of 
 the initial parameters of the order of $M$.

 In this paper, we invoke the idea of multidimensional gravity which is a widely used tool for 
 obtaining new theoretical results 
 \cite{Abbott:1984ba,Chaichian:2000az,Randall:1999vf,Brown:2013fba,Bronnikov:2009zza}. 
 The paper \cite{Krause:2000uj} uses warped geometry to solve the small cosmological constant 
 problem. Multidimensional inflation is discussed in 
 \cite{2002PhRvD..65j5022G,Bronnikov:2009ai,Fabris:2019ecx} where it was supposed that an 
 extra-dimensional metric $g_\n$ is stabilized at a high-energy scale. Stabilization of 
 extra space as a pure gravitational effect has been studied in 
 \cite{2003PhRvD..68d4010G,2002PhRvD..66d4014G}, see also \cite{Arbuzov:2021yai}.

 The present research is also based on nonlinear $f(R)$ gravity. The interest in $f(R)$ theories 
 is motivated by inflationary scenarios starting with Starobinsky's paper \cite{Starobinsky:1980te}. 
 At present, $f(R)$ gravity is widely discussed, leading to a variety of consequences, in particular, 
 the existence of dark matter \cite{Gani:2014lka,Arbuzova:2021etq}. Including a function of the 
 Ricci scalar, $f(R)$, is the simplest extension of general relativity. In the framework of such an 
 extension, many interesting results have been obtained. Some viable $f(R)$ models in 4D space 
 that satisfy the observational constraints are proposed in 
 \cite{DeFelice:2010aj,2014JCAP...01..008B,Sokolowski:2007rd,2007PhLB..651..224N,Nojiri_2017,}. 

 The idea that the Lagrangian parameters can be considered as some functions of a field has been 
 widely used since Schwinger's paper  \cite{Schwinger:1951nm,}. Such fields can be involved in the 
 classical equations of motion together with the ``main'' fields or treated as background fields. 
 The latter were applied for fermion localization on branes \cite{Sorkhi:2018nln,Sui:2017gyi,Arai:2018hao}, 
 gauge field localization \cite{Chumbes:2011zt}, extensions of gravity in a scalar-tensor form 
 (with $f(\phi)R$) \cite{Bronnikov:2003rf} and so on. In this paper, we show that a self-gravitating 
 scalar field can serve as a reason for the emergence of small parameters.

 As a mathematical tool, we use the Wilsonian approach technique, a well-known method for theoretical 
 studies of the energy dependence of physical parameters \cite{Peskin:1995ev}. In this approach, 
 the physical parameters $\lambda_i (M)$  of the Wilson action are fixed at a high energy scale $M$. 
 The renormalization flow used to descend to low energies (the top-down approach) is discussed in 
 \cite{Burgess:2013ara,Hertzberg:2015bta,Babic:2001vv,Dudas:2005gi,Wetterich:2001kra}. In particular, 
 quantum corrections to the Starobinsky model were discussed in \cite{RomeroCastellanos:2018inv}.

 In our approach, we add extra dimensions and study their role in different scales, with a hope that 
 it should make the renormalization procedure much more efficient. We make use of the idea of flexible 
 (inhomogeneous) extra space that has been developed in \cite{Gani:2014lka,Rubin:2015pqa,Rubin:2014ffa}. 
 Our preliminary study of inhomogeneous extra metrics concerns such parameters as the cosmological 
 constant \cite{Rubin:2015pqa}, those of the Starobinsky inflationary model and baryon asymmetry 
 of the Universe \cite{Nikulin:2020nub,Nikulin:2021bmw}. It has been shown there that inhomogeneous 
 metrics can be tuned to explain the smallness of the appropriate effective parameters.
 For example, encouraging results for explaining the smallness of the cosmological constant was 
 obtained in \cite{Rubin:2015pqa, 2017JCAP...10..001B}. The effect of quantum corrections in this 
 context was discussed in \cite{Rubin:2020pqu}. 

 Here we continue this research by including the Higgs sector of the Standard Model. There are three 
 energy scales which we intend to describe --- inflationary stage, the electroweak scale, and the 
 cosmological one. Each of them is characterized by a specific small parameter. The initial parameters 
 and the Lagrangian of our model are fixed at a sub-Planckian scale and do not vary during the Universe 
 evolution. Special attention is paid to the emergence mechanism small values, specific for each of 
 the three scales.

\section{The model}	

  Consider $f(R)$ gravity with a minimally coupled scalar field $\zeta$ in 
  a $\D = 4 + n$-dimensional manifold $M_\D$: 
\begin{eqnarray}\label{S0_}
	S = \frac{m_{\D}^{\D-2}}{2}  \int_{M_\D}  d^{\D} X \sqrt{|g_{\D}|} \,  \Bigl( f(R) 
	+ \partial^{\M}\zeta \, \partial_{\M}\zeta -2 V(\zeta) \Bigr)+S_{H_P}\, ,
\end{eqnarray}
 where $g_{\D} \equiv \det g_{\M\N}$, $\M,\N =\overline{1,\D}$, 
 the $n$-dimensional manifold $M_n$ is assumed to be closed, $f(R)$ is a function of the 
 D-dimensional Ricci scalar $R$, and $m_\D$ is the $\D$-dimensional Planck mass. Below, we 
 will work in the units $m_D=1$. The term $S_{H_P}$ denotes the Higgs action \eqref{SH} 
 considered in Sec.\,\ref{Higgs}, and it
 is assumed to be small as compared to the gravitational part of the action. 
 It is also postulated that the scalar field $\zeta$ is very massive and hence unobservable. 
 Nevertheless, this field plays a key role being responsible for the emergence of small 
 parameter(s), see a discussion at the beginning of Sec.\,\ref{intermed}.

 Variation of the action \eqref{S0_} with respect to the metric $g^{\M\N}_\D$ and the scalar 
 field leads to the known equations
\begin{align}         \label{eqMgravity}
    &-\frac{1}{2}{f}(R)\delta_{\N}^{\M} + \Bigl(R_{\N}^{\M} +\nabla^{\M}\nabla_{\N} 
        - \delta_{\N}^{\M} \Box_{\D} \Bigr) {f}_R  = - T_{\N}^{\M}, 
\\          \label{eqMscalarfield} 
	&\Box_{\D} \, \zeta + V_{\zeta} =0,  
\end{align}
  with $f_R = {df(R)}/{dR}$, $\Box_{\D}= \nabla^{\M} \nabla_{\M}$, and 
	$V_{\zeta} = {d V(\zeta)}/{d\zeta}$. 

 Equation \eqref{eqMscalarfield} is known to be a consequence of equations \eqref{eqMgravity}. The 
 corresponding stress-energy tensor of the scalar field $\zeta$ is
\beq
     T_{\N}^{\M} = \frac{\partial L_{\rm matter}}{\partial\bigl(\partial_{\M} \zeta \bigr)}
     \partial_{\N} \zeta - \frac{\delta_{\N}^{\M}}{2} L_{\rm matter}  
     = \partial^{\M}\zeta \, \partial_{\N}\zeta - \frac{\delta_{\N}^{\M}}{2} \, \partial^{\K}\zeta \, \partial_{\K}\zeta +  \delta_{\N}^{\M}V\bigl(\zeta \bigr) \, .
\eeq
Here the Higgs field contribution is omitted.
 We use the conventions for the curvature tensor $R_{\ \M\N\K}^\LL
 =\partial_\K\Gamma_{\M\N}^\LL-\partial_\N \Gamma_{\M\K}^\LL +\Gamma_{\CC\K}^\LL\Gamma_{\N\M}^\CC-\Gamma_{\CC\N}^\LL \Gamma_{\M\K}^\CC$ and the Ricci tensor $R_{\M\N}=R^\K_{\ \M\K\N}$. 

 The metric is supposed in the form
\begin{equation}\label{interval}
		ds^2=\e^{2 \gamma(u)}\left(dt^2 -\e^{2Ht}(dx^2 +dy^2 +dz^2)\right) 
			- du^2 -r(u)^2 d \Omega_{n-1}^2,
\end{equation}
 where $d \Omega_{n-1}^2$ is the metric on a unit $n-1$-dimensional sphere.
 The metric ansatz used in this paper has been widely studied in the framework of linear gravity \cite{2000PhRvD..62d4014O,2003PhRvD..68b5013C,2005PhRvD..71h4002S,2005PhRvD..71j4018R}, 
 applying, in particular, to solving the Hierarchy problem 
 \cite{2000PhRvL..84.2564G,2000PhRvL..85..240G,Bronnikov:2007kw}.
   
  The field equations for the metric \rf{interval} and $\zeta = \zeta(u)$ read
\begin{align}\label{tt}
& {R'}^2 f_{RRR} +\left[R'' +
\left(3 \gamma'  + (n-1) \frac{r'}{r}\right) R' \right]f_{RR} 
- \left( \gamma'' +4{\gamma'}^2 + (n-1)\frac{\gamma' r'}{r} 
- \frac{3H^2}{\e^{2 \gamma}} \right)  f_{R} \nonumber 
\\    &  \inch
- \dfrac{f(R)}{2} = - \dfrac{{\zeta'}^2}{2}   - V\bigl( \zeta \bigr), 
\\              \label{uu}
& \left( 4\gamma'R' + (n-1) \dfrac{r'}{r} R' \right)\,f_{RR} 
- \left( 4 \gamma'' + 4{\gamma'}^2 + (n-1) \dfrac{r''}{r} \right) \, f_R 
- \, \dfrac{f(R)}{2} =   \dfrac{{\zeta'}^2}{2}  - V\bigl(\zeta \bigr) \,, 
\\                  \label{aa}
		& {R'}^2 f_{RRR} \, + \bigg( R'' + 4\gamma' R' + (n - 2) \dfrac{r'}{r}\,R'\bigg) f_{RR} 
		 - \biggl(\dfrac{r''}{r} + \frac{4\gamma' r'}{r} + (n-2)\dfrac{{r'}^2}{r^2} 
		- \dfrac{(n-2)}{r^2} \biggr) f_R  
\nn  &   \inch
 		- \, \dfrac{f(R)}{2} = -  \dfrac{{\zeta'}^2}{2}  - V(\zeta) , 
\\      &            \label{scalar}
	\zeta'' + \left(4 \gamma' + (n-1) \dfrac{r'}{r}\right) \, \zeta'  - V_{\zeta} =0,
\end{align} 
  where the prime denotes $d/du$. Also, we will use the expression for the Ricci scalar
\beq      \label{Ricci_n}
        R(u)= \frac{12 H^2}{\e^{2 \gamma}} -8\gamma'' -20{\gamma'}^2 
        - (n-1) \left( \dfrac{2 r''}{r} + \frac{8\gamma' r'}{r}
        + (n-2) \left(\dfrac{r'}{r}\right)^2\,- \dfrac{(n-2)}{r^2} \right)
\eeq
 as an additional equation and $R(u)$ will be treated as a new unknown function to avoid 3rd and 4th order derivatives in \eqs \rf{tt}--\rf{aa}. 
 It can be shown that one of the equations \rf{tt}--\rf{scalar} is a consequence of the others. 
 The combination $2\times$\eqref{uu}$ - f_R \times $\eqref{Ricci_n}  is the constraint equation 
\bearr             \label{cons}
     \biggr(8\gamma' + 2 (n-1) \dfrac{r'}{r} \biggl) R'  f_{RR} + \Biggl(12 {\gamma'}^2
        + (n-1) \biggl(\dfrac{8\gamma' r'}{r} + (n-2)\dfrac{\bigl({r'}^2-1\bigr)}{r^2} \biggr) +R \Biggr)f_R
\nnn \inch
   - \frac{12 H^2}{\e^{- \gamma(u)}} f_R\, - f(R) 
          =  {\zeta'}^2 - 2 V\bigl( \zeta \bigr)	
\ear
  containing only first-order derivatives. It plays the role of a restriction on the solutions 
  of the coupled second-order differential equations \rf{tt}--\rf{Ricci_n}. 

  As a result, we use three independent equations \eqref{tt}, \eqref{aa}, \eqref{Ricci_n} and 
  the constraint \eqref{cons} to fix three functions $r(u),\gamma(u), R(u)$ and the unknown metric parameter $H$.
 
\begin{figure}[!th]
\bigskip
\centering
\includegraphics[width=0.2\linewidth]{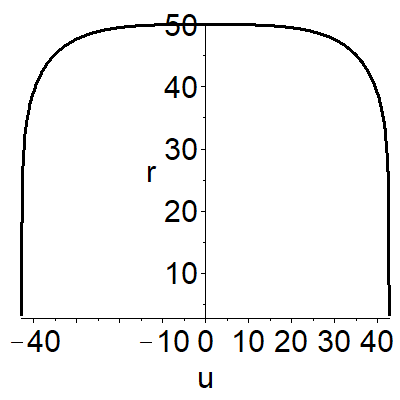} \quad 
\includegraphics[width=0.2\linewidth]{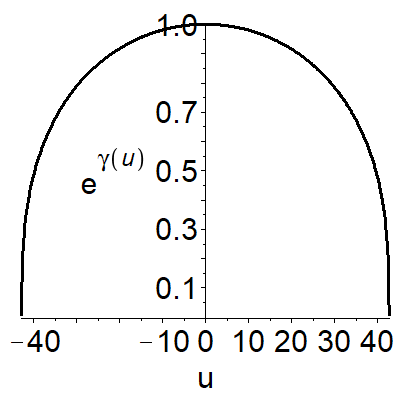}\quad 
\includegraphics[width=0.2\linewidth]{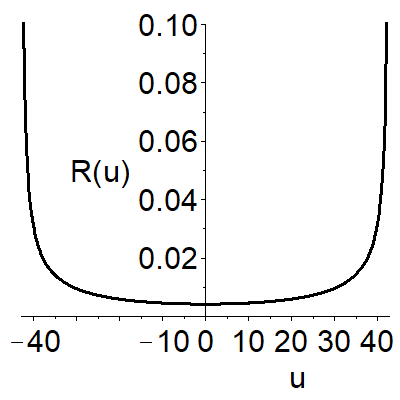}\quad 
\includegraphics[width=0.2\linewidth]{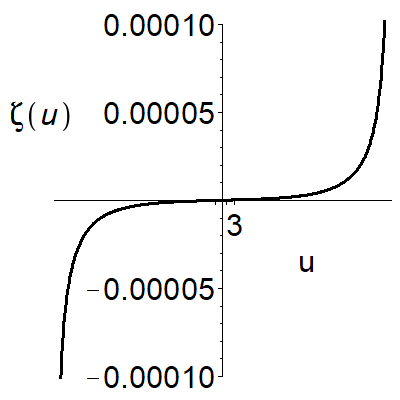}
\caption{\small
    Solution of \eqs \rf{tt}--\rf{Ricci_n} for the following parameters $n=3$, $f(R) =300 R^2 +R +0.002$,  $H = 0.003$, $V(\zeta)=0.01\,\zeta^2/2$ and boundary conditions  $r(0) =50$, $\gamma(0)=0$, $R(0) \simeq 0.00415$, $\zeta(0)=0$,  
    $r'(0) = \gamma'(0) = R'(0) = 0$, $\zeta'(0) = 6\times 10^{-8}$. }
\label{f1}
\end{figure}

  One of the possible numerical solutions to this system is shown in Fig.\,\ref{f1}. We note that the 
  warp factor $e^{\gamma(u)}\to 0$ at the boundaries, which are singular ends of the range of $u$
  and can be imagined as a kind of poles in a closed $n$-dimensional manifold since there $r\to 0$.
  The qualitative behavior of the solution shown in Fig.\,\ref{f1} is quite generic.
  The particular form of solutions, including the field distribution and the extra metric, 
  depends on the Lagrangian parameters postulated from the beginning. It also depends on the 
  boundary conditions at $u=0$ that are necessary for solving the second-order differential 
  equations. Unlike the Lagrangian parameters, the boundary conditions ultimately depend on random initial fluctuations within a pocket universe. Inflation produces a continuum set of such 
  universes with different initial conditions and therefore with different metric functions 
  and field distributions in the extra space.

\section{Matter localization around a singularity}

 In general, it is assumed here that matter is distributed throughout the extra 
 dimensions like in the Universal Extra Dimensional approach \cite{ArkaniHamed:1998rs,2003PhRvD..68f3516B}. 
 At the same time, there is another direction that deserves discussion. Indeed, we see from the 
 figures that there are two points where the metric is singular or has sharp peaks. They could 
 indicate the formation of branes if the extra space is large enough and if matter is 
 concentrated in a close neighborhood of these peaks (certainly assuming that the formal infinities are somehow suppressed by quantum effects). 
 This opportunity is briefly discussed in this section. 

 As a rough approximation, consider the motion of classical particles near such a singular point, 
 see Fig.\,1, bearing in mind the metric  
\beq        \label{n3}
    ds^2=\e^{2\gamma(u)}(dt^2 -dx^2 -dy^2 -dz^2) - du^2 -r(u)^2 (d\xi^2 + \sin^2 \xi\, d \psi^2).
\eeq
 The geodesic equations have the form 
\bearr        \label{tss}
        t_{ss} +2 \, t_s \, \gamma_u \, u_s =0,  
\yyy         \label{xss}
        x_{ss} +2 \, x_s \, \gamma_u \, u_s =0, \quad 
        y_{ss} +2 \, y_s \, \gamma_u \, u_s =0, \quad
        z_{ss} +2 \, z_s \, \gamma_u \, u_s =0, \quad
\yyy
        u_{ss} +\e^{2 \gamma} \, \gamma_u \,({t_s}^2 -{x_s}^2 -{y_s}^2 -{z_s}^2) 
        -r \, r_u \, {\xi_s}^2 -r \, r_u \sin^2 \xi\,{\psi_s}^2 =0,  
\yyy
        \xi_{ss} +2 \, \xi_s \, \frac{r_u}{r} \, u_s -\sin \xi\, \cos \xi \, {\psi_s}^2 =0,  
\yyy
        \psi_{ss} +2 \, \psi_s \, \frac{r_u}{r} \, u_s +2\cot \xi\, \xi_s \, \psi_s =0,  
\ear
where  the index $s$ denotes the derivative with respect to $s$, and the index $u$ denotes the derivative with respect to $u$. 
These equations admit solutions when $x, y, z, \xi$, and $\psi$ are constant. Let us assume, for simplicity, that $\xi=\pi/2$, then
\bearr
    0=t_{ss} +2\,t_s\,\gamma_u\,u_s = t_{ss}+2\,t_s\,\gamma_s \ \ \Rightarrow \ \ t_s =\e^{C_1 -2\gamma},
\yyy
    0 = u_{ss} +\e^{2 \gamma} \, \gamma_u \,{t_s}^2 
    = u_{ss} + \gamma_u \, \e^{2C_1 -2 \gamma} 
    = \frac{1}{u_s} \left(u_{ss}\,u_s + \gamma_s\,\e^{2C_1 -2\gamma} \right) 
    \ \ \Rightarrow\ \  {u_s}^2 = \e^{2C_1 -2 \gamma} +C_2,
\ear
 with integration constants $C_i$. The normalization relation gives
\beq
        1 = \e^{2 \gamma} {t_s}^2 -{u_s}^2 = -C_2.
\eeq
 Then
\begin{equation}
        {u_s}^2  =\e^{2C_1 -2 \gamma} -1 = {u_t}^2 \e^{2C_1 -4 \gamma}  
        \  \Rightarrow\   {u_t}^2 = \e^{2 \gamma}\left(1 -\e^{2\gamma - 2 C_1 }\right).
\end{equation}
\begin{figure}[!th]
\centering
\includegraphics[width=0.4\linewidth]{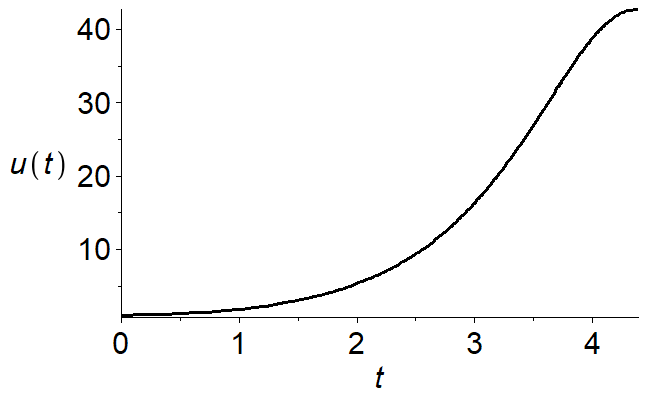} \qquad \includegraphics[width=0.4\linewidth]{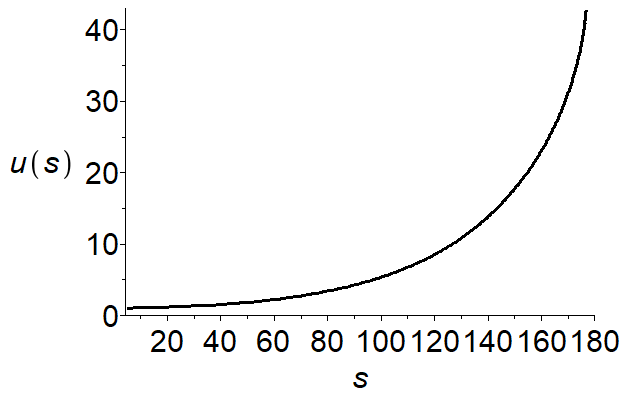}
\caption{$u\Big|_{t=0} = u\Big|_{s=0} = 1, \ du/dt\Big|_{t=0} = du/ds\Big|_{s=0} = 0$, in 
    the background of the solution shown in Fig.\ref{f1}. }
\label{f6}
\end{figure}

  For nonrelativistic particles, only the second equation matters. It can be approximated as
\begin{equation}
        u_{ss}\simeq -2e^{2\gamma}\gamma_u.
\end{equation}
  We see that the acceleration of a particle is directed to a singular point, which should 
  ultimately lead to concentration of matter at such a point.
  As a result, matter is localized around both ``poles,'' as should be the case in a brane world
  (this time consisting of two branes on the two ``poles'').
  It opens a door for developing a mechanism of strong reduction of the initial parameter values. 
  For example, an interaction term of the form
\[
    \kappa \int d^D Z\sqrt{|g_D|} \chi(z)\bar{\psi}(z)\psi(z)
\]
  contains overlapping integral
\[
    I_{\rm overlap}\equiv \int d^n y \sqrt{|g_n|}\chi(y)\bar{\psi}(y)\psi(y)
\]    
  over the extra dimensions which could be arbitrarily small if the fields $\chi(y)$ and 
  $\psi(y)$ are localized near different branes. It leads to the coupling constant renormalization
\[
        \kappa\to \kappa' =\kappa I_{\rm overlap}\ll \kappa.
\]        
  We will leave this idea for future studies and return to our main discussion.

\section{Intermediate energies. The Starobinsky model}\label{intermed}

 The second energy scale relates to the inflationary stage with the small parameter 
 $H/m_{\Pl}\sim 10^{-6}$. Different inflationary models use different parameters of this order. 
 It could be the inflaton mass in the simplest model of inflation with a quadratic potential 
 or a constant factor of $R^2$ term in the Starobinsky model.
 Let us restore the latter. To this end, we should solve the system 
 \eqref{tt}--\eqref{scalar} and obtain the necessary values of its parameters. 

 The scalar field $\zeta$ affects the extra-space metric through the Einstein equations, but here 
 we are interested in small amplitude solutions of this field, i.e., $\zeta(X)\ll 1$.
 Therefore, its role in the metric formation is negligible, and it can be considered as a test field 
 acting in the background metric. This approximation makes the analysis easier but is not very 
 significant for our reasoning. 

\subsubsection*{The emergence of small parameters}


 A successful solution of the Hierarchy problem implies the presence of small parameters, and we have 
 enough tools to create them. Indeed, in our picture, there is an infinite set of different universes 
 created during inflation \cite{Guth:1980zm,Lindebook,Fabris:2019ecx} which contain the independently 
 fluctuating field $\zeta$. These fluctuations decay with time and lead to static field distributions 
 in each universe. There is an infinite set $\aleph$ of such static distributions 
 that form a continuum set. The situation is similar to the boson star formation 
 model \cite{Liebling:2012fv} where a self-gravitating scalar field forms a variety of dense 
 stable clumps. The set $\aleph$ contains a subset of small-amplitude distributions $\zeta(u)$ like those presented 
 in Fig.\,\ref{f1}, right panel. Their values averaged over the extra dimensions represent a set 
 of small parameters to be widely used below.

 \bigskip

 To proceed, let us restore some formulas from our previous paper \cite{Petriakova:2023klf} for a
 relation between the $\D$-dimensional Planck mass and the 4-dimensional one, which is needed to 
 convert units $m_\D=1$ into the physical units. To this end, define
\begin{equation} \label{R4}
        R_4 \equiv 12 H^2, \qquad R_n \equiv R(u) - \e^{-2 \gamma(u)} R_4,
\end{equation}
  see \eqref{Ricci_n}. Substitution of the Taylor series
\beq \label{ser}
        f(R) \simeq f(R_n) + f_{R}(R_n)\e^{-2\gamma(u)}R_4 
         + \frac{1}{2}f_{RR}(R_n)\e^{-4\gamma(u)} R_4^2 + \ldots
\eeq
 into the gravitational part of the action \eqref{S0_} leads to an effective theory after 
 integration over the extra coordinates:
\begin{equation}\label{S_eff}
	S_{\eff} = \dfrac{m_{\Pl}^{2}}{2} \int\limits_{M_4} d^4 x \sqrt{|g_4|} 
        \Bigl(a_{\eff} R_4^2 + R_4 + c_{\eff} \Bigr). 
\end{equation}
  Here $g_4$ is the determinant of the 4D metric
\begin{equation}\label{g4}
        ds^2 = g_{\mu\nu} dx^\mu dx^\nu = dt^2 - \e^{2Ht}\delta_{ij}dx^i dx^j \, ,
\end{equation}
  and
\bearr          \label{mPl}
        m^2_{\Pl}  =  \mathcal{V}_{n-1} \int_{u_{\min}}^{u_{\max}}
                    f_R (R_n) \,\e^{2\gamma}\,r^{n-1}\, du,
\yyy               \label{a_eff}
        a_{\eff}  = \frac{\mathcal{V}_{\n-1}}{2m_{\Pl}^{2}} 
                \int_{u_{\min}}^{u_{\max}} f_{RR}(R_n) \,\e^{4\gamma}\,r^{n-1}\,du,
\yyy\label{c_eff}
        c_{\eff}  =  \frac{\mathcal{V}_{n-1}}{m_{\Pl}^2} \int_{u_{\min}}^{u_{\max}} 
                    \Bigl(f(R_n) - (\zeta')^2  - 2V(\zeta)\Bigr)\,\e^{4\gamma}\,r^{n-1}\, du .
\ear
 where $\mathcal{V}_{n-1} = \int d^{n-1} x \sqrt{|{g}_{n-1}|} = \dfrac{2\pi^{n/2}}{\Gamma(n/2)}$. 
 The r.h.s. of \eq \eqref{mPl} is written in units $m_D=1$. This relation is used to express 
 the D-dimensional Planck mass in terms of the 4D Planck mass $m_{\Pl}$. Here we suppose that
 the functions $\gamma(u),\ r(u),\ \zeta(u),\ R(u)$ form a particular solution 
 to the system \eqref{tt}-\eqref{scalar} for a specific value of $H$. {Therefore the values of
 $a_{\eff}(H)$ and $c_{\eff}(H)$ are functions of the Hubble parameter. They are approximately 
 constants during inflation and at the present times, being different in these two periods. The 
 parameter $ c_{\eff}(H)$ is fixed at the present epoch when $H\lll m_{\Pl}$, while 
 the parameter $a_{\eff}(H)$ is determined by the appropriate inflation rate.}
 The parameter $a_{\eff}$ must be approximately equal to the observable value obtained from the 
 COBE normalization \cite{Planck:2018jri}, as 
\begin{equation}         \label{a}
        a_{\text{Starob}} \simeq 1.12 \cdot 10^9 \left(\dfrac{N_\e}{60}\right)^2m^{-2}_{\Pl}. 
\end{equation}
  For the solution shown in Fig.\,\ref{f1}, $a_{\eff} \simeq 7.2 \cdot 10^{8}\, m_{\Pl}^{-2}$, 
  and the Hubble parameter is $H\simeq 1 \cdot 10^{-6} m_{\Pl}$. One can see that the 
  Starobinsky inflationary model has been restored. The parameter values $a=300,\, c=0.002 $  
  lead to the following values of the dimensionless parameters:
\begin{equation}\label{acp}
    a'=\sqrt{am_D^2}\simeq 17,\qquad c'=\sqrt{c/ m_D^2}\simeq 0.045
\end{equation} which looks natural.
  
  The other parameter, $c_{\eff}(H)$, needs a separate discussion. Equation \eqref{c_eff} for  
  $c_{\eff}(H)$ is derived under the assumption that all functions in \eqs \eqref{tt}--\eqref{scalar} 
  are stationary, which takes place for a 4D de Sitter metric in which the Hubble parameter $H=\const$. 
  This approximation is valid at slow-roll inflation with the small parameter $|\Dot{H}|/H^2 \ll 1$. 
  Fortunately, this inequality holds for a wide range of the parameter $c_{\eff}$, in particular 
  for $c_{\eff}=0$ \cite{Starobinsky:1980te}. However, \eq \eqref{c_eff} has a practical 
  meaning only for a pure de Sitter metric but not for slow rolling. This formula could be valid at the present times, for example, if we suppose that the cosmological constant $\Lambda=-c_{\eff}/2$ is 
  really constant. It is a subject of detailed discussion, see Sec.\,\ref{CC}. 


\section{The electroweak scale. Restoration of the Higgs parameters}\label{Higgs}

  In this section, the reasoning is in the spirit of our previous paper \cite{Petriakova:2023klf}, 
  but without introducing an external scalar field.

\subsection{Analytical formulas}

  In the previous section, we have reproduced the Starobinsky model of inflation at the scale 
  of $10^{13}$\,GeV. The appropriate values of the initial Lagrangian parameters $a, c$ are fixed. 
  These parameters must be the same at low scales where the Hubble parameter is equal to zero, 
  $H\simeq 0$, as compared to the Planck scale. On the contrary, the extra space metric depends 
  on the energy scale, the Hubble parameter in our case.

  In this section we discuss the Hierarchy problem at the electroweak scale using the Higgs 
  field as an example. Within 
  the framework of our approach outlined in the Introduction, we assume that the physics of the 
  Higgs field is formed at the Planck scale. 

  Suppose that the form of the Higgs action at the Planck scale is the same as at the electroweak scale, 
\beq            \label{SH}
    S_{\HH_\text{P}} =
        \frac12 \int d^{\D} X \sqrt{|g_{\D}|} \, 
        \Bigl(\partial^{\M} {H_\text{P}}^\dagger \partial_{\M} H_\text{P} 
    + \nu {H_\text{P}}^\dagger H_\text{P} - \lambda\bigl({H_\text{P}}^\dagger H_\text{P}\bigr)^2 \Bigr), 
\eeq
  where the symbol $\dagger$ means Hermitian conjugation,
  $\nu$ and $\lambda  > 0$ are arbitrary numbers and $H_\text{P}$ is a proto-Higgs field. 

  We have managed to avoid large/small initial dimensionless parameter values $a',c'$, see \eqref{acp} when describing the Starobinsky model acting 
  at energies $\sim 10^{13}$\,GeV. Our intention is to repeat 
  this success at the electroweak energies. To do that, we need to show that the initial parameters
  can be reduced by many orders of magnitude. All numerical values in the Lagrangian \eqref{SH} are 
  of the order of unity in $m_\D$ units. More definitely, let us express the dimensionful parameters 
  $\nu, \lambda$ in terms of the dimensionless ones $\nu', \lambda'$:  
\begin{equation}            \label{dimless}
        \nu\to (\nu' m_D)^2,\qquad      \lambda \to (\lambda'/m_D)^{D-4}.   
\end{equation} 
  It is these dimensionless parameters $\nu'$ and $\lambda'$ that should vary around unity.

  The classical equations of motion are obtained by varying the action \eqref{SH} with respect to $H_\text{P}$, which gives
\begin{equation}            \label{boxHP}
        \square_\D H_\text{P}=\nu H_\text{P} 
        - 2 \lambda  \bigl({H_\text{P}}^\dagger H_\text{P}\bigr)H_\text{P}.
\end{equation}
  The proto-Higgs field can be presented as
\begin{equation}
        H_\text{P} = h(x) \ {U}(u) + \delta H_\text{P}, \qquad 
        \delta H_\text{P} = \sum_{k} h_k(x) Y_k(u)
\end{equation}
  where $h(x)$ and $h_k(x)$ are 2-component columns acting in the fundamental representation of $SU(2)$. 
  In what follows, we will consider the case 
\begin{equation}            \label{Hxu}
        H_\text{P} \simeq h(x)\,U(u),\qquad \delta H_\text{P}\ll h(x)\,U(u).
\end{equation}
  The dimensionality of the proto-Higgs field is $[H_P]=m_D^{(D-2)/2}$, $[h]=[h_k]=m_D$, 
  $[U]=[Y_k] = m_D^{n/2}$.

  Our immediate aim is to find the distribution of the field $H_\text{P}$ over the extra coordinates 
  governed by the scalar function $U(u)$ by solving \eqs \eqref{boxHP}, \eqref{Hxu}.
  The inhomogeneities of the field $h(x)$ are important at low energies, but they are exponentially stretched during the first de Sitter-like stage so that $h(x)= \const$ 
  with great accuracy. It means that 
\begin{equation}            \label{Hv}
h(x) = \frac{1}{\sqrt{2}} 
\begin{pmatrix}
0 \\ v_0+\rho(x)
\end{pmatrix}
\simeq 
\frac{1}{\sqrt{2}} 
\begin{pmatrix}
0 \\ v
\end{pmatrix}.
\end{equation}
  Therefore, the approximation \eqref{Hv} transforms \eq\eqref{boxHP} in the following way:
\begin{equation}        \label{boxU}
        \square_n U(u) =\nu U(u) - \lambda\,v^2 U^3(u), 
\end{equation}
  with a yet unknown parameter $v$. We suppose further on that the metric functions as well as the Lagrangian parameters remains the same as those considered above with one 
  exception: the Hubble parameter is extremely small at the present epoch as compared to the 
  inflationary epoch, and below we put $H\approx 0$.

  The knowledge of solutions to \eq \eqref{boxU} permits us to integrate out the internal coordinates 
  and to reduce the action \eqref{SH} to the 4D form
\bearr           \label{SHH}
    S_{\HH} = \frac{\mathcal{V}_{\n-1}}{2}\int d^{4} x \sqrt{|\tilde{g}_{4}|} 
                \int_{u_{\min}}^{u_{\max}}\biggl[\e^{-2\gamma(u)}  U^2(u) \tilde{g}^{i j} \partial_{i} h^\dagger\partial_{j} h 
\nnn \inch
        + \Bigl(-(\partial_u U)^2+ \nu\, U^2(u)\Bigr) h^\dagger h - \lambda\, U^4(u) 
        \bigl( h^\dagger h\bigr)^2 \biggr] \e^{4\gamma(u)} r^{n-1}(u)\, du
\earn
  after substitution of \eqref{Hxu} into \eqref{SH}. 
  To study this action at low energies, we choose the Minkowski metric
\begin{equation}
        \tilde{g}_{4,ij} = \eta_{i j}     
\end{equation}
  and define the following parameters by integration over $u$:
\bearr           \label{Kh}
        K_h = \mathcal{V}_{\n-1} \int_{u_{\min}}^{u_{\max}} U^2(u)\,\e^{2\gamma(u)}r^{\n-1}(u)\,du,  
\yyy            \label{m_rho}
        m^2_h =\mathcal{V}_{\n-1} \int_{u_{\min}}^{u_{\max}}
        \Bigl(-(\partial_u U)^2 +\nu\, U^2(u)\Bigr)\e^{4\gamma(u)}r^{\n-1}(u)\, du,  
\yyy             \label{lh}
        \lambda_h = \mathcal{V}_{\n-1} \int_{u_{\min}}^{u_{\max}}
        \lambda\,U^4(u)\,\e^{4\gamma(u)}r^{\n-1}(u)\, du.
\ear

\subsection{Comparison with the Higgs parameters}

  Recall that a natural range for the dimensionless parameters $\lambda'$ and $\nu'$ is 
  $10^{-2}$ to $10^2$. It means that acceptable ranges of the ``physical'' parameters are 
  $(10^{-6}\div 10^6)$ for $\lambda$ and $(10^{-4}\div 10^4)$ for $\nu$ according to 
  the definitions \eqref{dimless}.

  The substitution
\begin{equation}\label{Hh}
            H_0(x)={h(x)\sqrt{K_h}}
\end{equation}
  leads to the 4D effective Higgs Lagrangian
\bearr     \label{SHHH}
        S_{\HH} = \frac{1}{2}\int d^{4} x  \sqrt{|\tilde{g}_{4}|} 
        \biggl(\partial_{i} H_0^\dagger\partial^{i} H_0 
        +  m_H^2 H_0^\dagger H_0 -  \lambda_H\bigl(H_0^\dagger H_0\bigr)^2 \biggr)\, , 
\yyy  \cm 
        m_H^2 \equiv \frac{m_h^2}{K_h},\qquad \lambda_H \equiv \frac{\lambda_h}{K_h^2}.
        \label{mlH}
\ear
  Here $H_0$ is the observable Higgs field at zero energy. The experimentally measured 
  parameters are the Higgs boson mass and its vacuum average,
\begin{equation}\label{obs}
        m_{\rm Higgs}=125 \,\text{GeV},\quad v_{\rm Higgs}=246\,\text{GeV}\,.
\end{equation}
  according to \cite{Workman:2022ynf}. They are related to the parameters $m_H$ and 
  $\lambda_H$ of the effective Higgs action \eqref{SHHH} as follows: 
\begin{equation}            \label{mH}
        m_{H} =m_{\rm Higgs} /\sqrt{2}=88.6\,\text{GeV} \simeq 10^{-17} m_{\Pl}, 
\end{equation}
  and 
\begin{equation}\label{lH}
        \lambda_{H}=({m_{H}}/v_{\rm Higgs})^2/2  \simeq 0.13, 
\end{equation} 
  The vacuum energy of the Higgs field is
\[
        V_{\min} = -\frac12 m_H^2v_{\rm Higgs},
\]
  so that the parameter $c$ in the function $f(R)$ should be corrected, $c\to c+V_{\min}$. 
  Note, however, that $V_{\min}$ is very small as compared to the D-dimensional Planck scale and 
  may be neglected.

  The above formulas contain the function $U(u)$, the solution to \eq \eqref{boxU} with a yet 
  unknown constant $v$. It is of interest that the Lagrangian structure \eqref{SH} allows us to avoid 
  the determination of this constant. Indeed, a solution to \eq \eqref{boxU} can be found for the 
  function 
\[
        \tilde{U}(u) = vU(u)
\]    
  because \eq \eqref{boxU} $\tilde{U}(u)$ does not contain the unknown parameter $v$ in this case. 
  Moreover, substitution of $U=\tilde{U}(u)/v$ into the expressions \eqref{Kh}, \eqref{m_rho} and 
  \eqref{lh} gives
\begin{equation}\label{UtU}
        K_h[U]= \frac{K_h[\tilde{U}]}{v^2} , \qquad 
        m_h^2[U]=  \frac{m_h^2[\tilde{U}]}{v^2}, \qquad 
        \lambda_h[U]=  \frac{\lambda_h[\tilde{U}]}{v^4},
\end{equation}
  hence the observable parameters $m_H$ and $\lambda_H$ in \eqref{mlH} do not depend on $v$. 
  This quantity also appears in the relation 
\begin{equation}\label{vcalc}
        v \simeq v_{\rm Higgs}/\sqrt{K_h[U]},
\end{equation}
  following from \eqref{Hh} and the substitution $H_0\to v_{\rm Higgs},\, h\to v$.
  Luckily, this relation does not depend on $v$ as well. After taking into account the first equality 
  in \eqref{UtU}, we obtain an additional restriction for the function $\tilde{U}$:
\begin{equation}\label{vcalc1}
        1 \simeq v_{\rm Higgs}/\sqrt{K_h[\tilde{U}]}.
\end{equation}
  The quantity $K_h[\tilde{U}]$ is calculated in $m_D$ units. Therefore, $v_{\rm Higgs}$ should 
  be also expressed in $m_D$ units.
\begin{figure}[!th]
\centering
\includegraphics[width=0.38\linewidth]{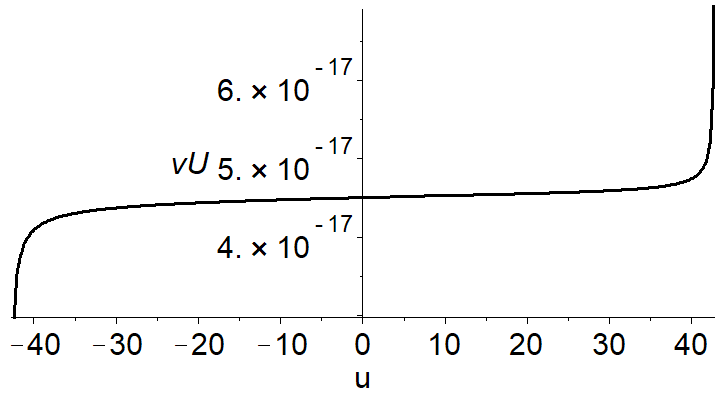}
\caption{\small
    Solution of \eq \eqref{boxU} for the function $\tilde{U}(u)=vU(u)$. Parameter values: 
    $\nu =7.5 \times 10^{-6}$,  $\lambda = 2.9\times 10^{5}$ (the dimensionless parameters defined 
    by \eqref{dimless} are as follows: $\nu'\simeq 0.003,\ \lambda'\simeq 70 $). Additional 
    conditions: $\tilde{U}(0) =4.5 \times 10^{-17}$, $\tilde{U}'(0) =2.6 \times 10^{-20}$ in the background shown in 
    Fig.\,\ref{f1}. For this solution, $v_{\rm Higgs}/\sqrt{K_h[\tilde{U}]} \simeq 0.99,\ 
    \lambda_H \simeq 0.13,\ m_H \simeq 10^{-17} m_{\Pl}$. }
\label{tildeU}
\end{figure}
  Figure\,\ref{tildeU} presents the Higgs field distribution $\Tilde{U}$ in the extra dimensions.

  In this section, we have found the conditions at which the initial parameter values written in Fig.\ref{tildeU} reproduce the Higgs Lagrangian with the observed parameters.
  The origin of small parameters has been discussed earlier, see the beginning of Sec.\,\ref{intermed}. 
  Quantum fluctuations produce a variety of field amplitudes in the countable set of pocket 
  universes. A small measure of them contains (extremely) small amplitudes $\Tilde{U}$.
  The fields values of the order of $\tilde{U}\sim 10^{-16}$ (see Fig.\,\ref{tildeU}) are
  suitable for the relations (\ref{mH}), (\ref{lH}) and (\ref{vcalc1}). 

\section{Low energies. The cosmological constant}\label{CC}

  Analytical formulas for 4D gravity have been obtained in Sec.\,\ref{intermed}. It is assumed that they 
  are approximately valid at the inflationary scale $\sim 10^{13}$\,GeV. The parameter $a_{\eff}$ was 
  obtained for specific values of the initial parameters $a, c, m_D$. A calculation of $c_{\eff}$ 
  representing the effective cosmological constant (CC) is not necessary because its value could vary 
  in a wide range without any effect on the inflationary process. This value is important at low 
  energies where the Hubble parameter is $\sim 10^{-61}\approx 0$\,{GeV$^2$}. 
  
  Therefore, the extra metric must be found by solving the Einstein equations at $H=0$. 
  Luckily, the resulting metric weakly depends 
  on the Hubble parameter if it varies within the interval $0 < H < 0.01$, and Fig.\,\ref{f1} gives 
  the appropriate impression. 

  At the low energy scale, the Hubble parameter $H$ is small, and the curvature squared $R_4^2$ can 
  be neglected in \eqref{S_eff}. In this case, the 4D Einstein equations lead to a relation between the
  CC and the Hubble parameter,
\begin{equation}     \label{LH}
        c_{\eff}\equiv -2\Lambda  = -6\, H^2,
\end{equation}
  which means that $c_{\eff}$ should be extremely small as well.

\begin{figure}[!th]
\centering
\includegraphics[width=0.45\linewidth]{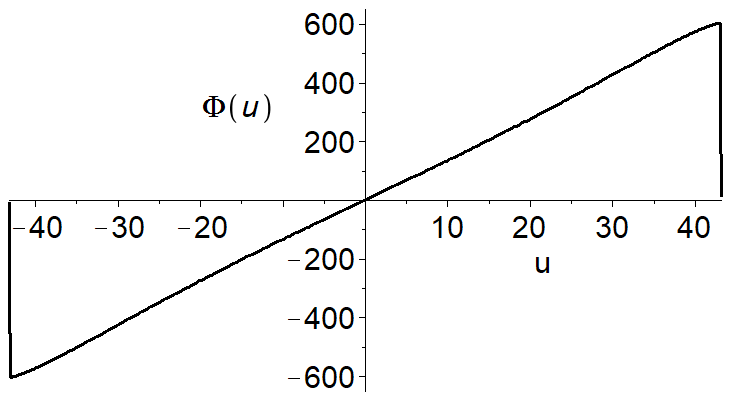} \quad \includegraphics[width=0.35\linewidth]{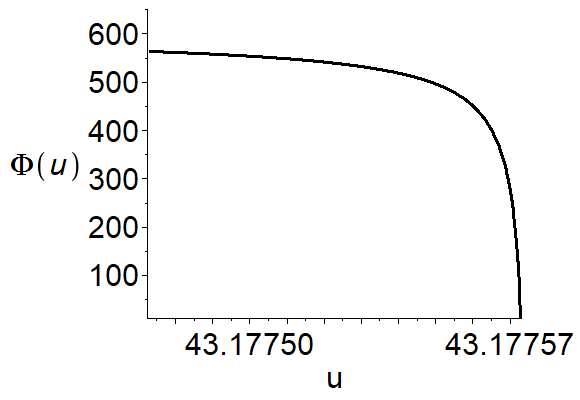}
\caption{\small
The integrand $\Phi (u)= \Big(f_{RR}\,  R' -f_R\, \gamma' \Big) \,\e^{4\gamma} r^{\n-1}$ in the right-hand side of \eq \rf{ceff_2} is zero at the boundary points 
$u_{\max}$ and $u_{\min}$.
Solution of equations Eqs.\eqref{tt}--\eqref{Ricci_n} is taken for parameters $n=3$, $f(R) =300 R^2 +R +0.002$,  $H = 0$, $V(\zeta)=0.01\,\zeta^2/2$ and boundary conditions  $r(0) =50$, $\gamma(0)=0$, $R(0) \simeq 0.00396$, $\zeta(0)=0$,     $r'(0) = \gamma'(0) = R'(0) = 0$, $\zeta'(0) = 3.62\times 10^{-5}$.}
\label{int}
\end{figure}

  On the other hand, the same value was found above starting from the initial D-dimensional action, 
  see \eq \eqref{c_eff}. It can be presented in the following form (see the Appendix):
\begin{eqnarray}\label{ceff_2}
    c_{\eff} &=& - 6 H^2 + \mathcal{V}_{\n-1} \frac{m_{\D}^{\D-2}}{m_{\Pl}^{2}} 2\int_{u_{\min}}^{u_{\max}} 
    \Big[ \Big(f_{RR}\,R'- f_R\,\gamma'\Big)\,\e^{4\gamma} r^{\n-1} \Big]'\, du + O(H^6).
\end{eqnarray}
  A comparison of the expressions \eqref{LH} and \eqref{ceff_2} derived with arbitrary initial 
  parameters and boundary conditions indicates that the integral in \eqref{ceff_2} must be zero. 
  It makes sense to prove this statement directly. To this end, we should find the function 
  $\Phi (u) \equiv \Big(f_{RR}\,  R' -f_R\, \gamma' \Big) \,\e^{4\gamma} r^{\n-1}$ 
  at the boundary points $u_{\min}$ and $u_{\max}$.
  Numerical simulations indicate that this function tends to zero indeed, see Fig.\ref{int}. Unfortunately, the accuracy 
  is unsatisfactory while approaching the boundary points. To clarify the situation, the following 
  can be suggested: we modify the nonlinear term in the action $R^2\to R^2 e^{-\epsilon R^2}$ from 
  the beginning, $\epsilon \lll 1$ in $m_D$ units, and put $\epsilon = 0$ finally. It does not 
  affect the equation of motion, but smooths out the singularities. In this case, $\Phi\to 0$ at the boundary points, where $r=0$ by definition. Hence, the integral as a whole equals zero.

  In this subsection, we have proved that the well-known relation $H^2=\Lambda/3$ can be derived from 
  D-dimennsional gravity. In the standard notations, $c_{\eff}=-2\Lambda$, where $\Lambda$ is the 
  cosmological constant. {In a more general case of the     
  the function $f(R)$, terms proportional to $H^6$ and other nontrivial terms can appear in 
  the expression \eqref{ceff_2}. In this case, the  inflationary dynamic requires a separate 
  study. 

\section{The role of quantum fluctuations}
\subsection{The smallest scale of compact extra dimensions}

  It is known that our instruments ``feel'' average values of a field $\Bar{\zeta}(u)$ calculated as 
\[
        \Bar{\zeta}(u) = \zeta_{\rm classical}(u) + \delta\zeta(u),
\]        
  where  $\zeta_{classical}(u)$ is the classical part, and $\delta\zeta(u)$ is a quantum correction to it. It makes sense to calculate the classical part only if $\zeta_{\rm classical}(u) \gg \delta\zeta(u)$.
  This inequality holds only if the action $S \gg 1$ 
  (the steepest descend method). Thus the classical approach can be applied in our approach if
\begin{equation}
        S = \int dv_D f(R) \gg 1,
\end{equation}
  or in other words
\begin{equation}
    S \simeq \delta v_D \langle f(R) \rangle \gg 1,
\end{equation}
  where $\delta v_D \simeq \delta u^D$ is a small volume parametrized by the coordinate $u$, 
  and $\langle ...\rangle$ stands for averaging over this volume. 
  It means that the volume $\delta v_D$ must not be smaller than 
  $\sim 1/\langle f(R)\rangle \to  \delta u \sim \langle f(R)\rangle^{-1/D}.$
  Thus for a classical description to make sense (i.e. to approximately coincide with the averages),
  the averaging range must not be smaller than 
\begin{equation}
        \delta u \sim  \langle f(R)\rangle^{-1/D}.
\end{equation}
  For example, for a seven-dimensional space and $f(R)\sim 10$, we have $\delta u \sim 10^{-1/7}\sim 1$.
  It means that the size of extra dimensions should be larger than $1/m_D$. Also, it is dangerous 
  to make physical conclusions based on classical solutions in the vicinity of singular points at
  a distance smaller than $1/m_D$.   

\subsection{Fluctuations in the present epoch}

  The field value $H_P\propto 10^{-17}$ is quite a small value. Let us estimate the probability of 
  large fluctuations that could destroy the solution.
  The cosmological probability to find a field value $\chi_2$ at the instant $t_2 = t_1 + t$ in 
  a spatial region of the horizon size $H^{-1}$ was studied in \cite{KhlopovRubin, 2021arXiv210908373R}. Based on the outcomes, it is possible to demonstrate that the first mode amplitude fluctuation probability, $h_1$, can be determined as
\begin{eqnarray}\label{Probc}
        dP = dP_1 = dh_1 \cdot\sqrt{q_1/\pi}\exp [-q_1\,h_1^2],\quad t\to \infty
\end{eqnarray}
  where
\begin{eqnarray}\label{key}
        q_1 = \frac{\mu}{\sigma^2}, \qquad\mu=\frac{m_1^2}{3H}, \qquad \sigma = \frac{H^{3/2}}{2\pi}. 
\end{eqnarray}
  $m_1 \sim m_D$, and the present-day Hubble parameter is $H=1.2\times 10^{-61} m_{\Pl}$. 
  Their knowledge allows one to estimate the parameter $q_1$.
  The fluctuation $h_1$ should be of the order of the classical part, 
  $h_1\sim  \Tilde{U}\sim 10^{-17}$ (see Fig.\,\ref{tildeU}) or larger to destroy it. 
  Now we have everything to estimate the exponent, 
\[
        q_1\,h_1^2 \sim \e^{211} \left(\frac{m_D}{M_{\Pl}}\right)^4,\qquad  m_D>10^{-5}M_{\Pl}.
\]
This estimate can be substituted to \eqref{Probc} to demonstrate how unlikely it is 
    that even such tiny classical field ($\Tilde{U}\sim 10^{-17}$) will be destroyed.

\subsection{Quantum corrections}

  The essence of the Wilson approach is to fix a Lagrangian and its parameters at the highest scale 
  and shift down to a low energy scale. It is achieved by sequentially integrating the Euclidean 
  action over a small slice of the momentum interval $\Delta k_E$. The renormalization group equations 
  thus obtained are widely used in this concern \cite{Peskin:1995ev}. The relations between low-energy 
  parameter values and high-energy ones are discussed in \cite{Marian:2021tje}. Also, quantum 
  fluctuations could modify the same form of a Lagrangian, \cite{2018PhRvD..98d3505L,2022JCAP...03..058I}. 

  The inclusion of a compact extra space into consideration complicates the procedure. Indeed, we 
  cannot choose an arbitrarily small momentum interval due to the energy level discreteness. 
  For example, if a size is quite small, $\Delta k_E < 1/r$, $r$ being the scale of extra dimensions, 
  then this momentum interval does not contain energy levels at all. A possible way to overcome 
  this difficulty is discussed in \cite{Rubin:2020pqu}, where truncated Green functions
\[
        G_T(Z,Z')\equiv \sum_{N\in \cal{N}}\frac{Y_N(Z)Y_N(Z')^*}{\lambda_N}
\]
  were introduced. Here $Y_N(Z)$ is a subset of $n+4$-dimensional eigenfunctions. The coordinates 
  $Z$ describe both 4D space and a compact extra space. It allows for approximately calculating 
  the parameters at low energies. As a result, quantum corrections caused by a scalar field are proportional to its self-coupling. This means that such quantum effects cannot be responsible for reducing the parameter values by many orders of magnitude, from the Planck scale to the electroweak scale. The classical mechanism discussed in this paper was elaborated just for this aim.
  The procedure of quantum renormalization is a necessary and unavoidable element that leads to 
  fine tuning of the physical parameters at low energies. 


\section{Conclusion}

  In this paper, we have discussed an approach which provided the hierarchy of three energy scales, the inflationary, electroweak and cosmological ones. Necessary tools for small parameters formation and 
 a successful solution of the problem are $f(R)$ gravity and inhomogeneous extra dimensions.

  The set of small parameters is formed in the following way. Slow rolling of a spatial domain from 
  a sub-Planckian scale down to the inflationary one gives rise to several consequences: (1) nucleation 
  of an infinite set of causally disconnected domains (pocket universes), (2) quantum fluctuations in 
  each domain produce a variety of fields and an extra-space metric distribution, (3) these 
  distributions are stabilized when the energy scale is low enough. Self-gravitating (scalar) fields 
  do not necessarily settle at states with minimum energy. On the contrary, e.g., the boson stars 
  activity \cite{Liebling:2012fv} is based on the fact that self-gravitating scalar fields can settle 
  at a continuum set of static states. There are states with arbitrarily small amplitudes among them. 
  These states are formed in a small but finite set of universes. As a result, a small but nonzero 
  measure of different universes contains small effective parameters that are applied here to 
  solve the Hierarchy problem at three energy scales.

  The mechanism developed should be accompanied by a renormalization group analysis aimed at 
  correction of the initial parameter values.

\section*{Acknowledgements}
The work of SGR and KAB was funded by the Ministry of Science and Higher Education of the Russian Federation, Project "New Phenomena in Particle Physics and the Early Universe" FSWU-2023-0073
and the Kazan Federal University Strategic Academic Leadership Program. The work of AAP was funded by the development program of Volga Region Mathematical Center (agreement No. 075-02-2023-944).
KAB also acknowledges support from Project No. FSSF-2023-0003.

\section*{Appendix}

  The validity Condition of \eq \eqref{LH} is not so trivial in the $4+n$-dimensional case,
  and we will discuss it here.

  We can exclude the terms with a scalar field in the definition of $c_{\eff}$; using 
  the expression \eqref{tt}, we obtain
\begin{align}\label{intL_}
  c_{\eff} & = \mathcal{V}_{\n-1} \frac{m_{\D}^{D-2}}{m_{\Pl}^{2}} \int\limits_{u_{\text{min}}}^{u_{\text{max}}} \Bigl(f\bigl(R_{\n}\bigr) - \bigl(\zeta^{\prime}\bigr)^2  - 2V\bigl( \zeta \bigr)\Bigr) \, \e^{4\gamma}\,  r^{\n-1} \, du  
\nonumber \\
& =\mathcal{V}_{\n-1} \frac{m_{\D}^{D-2}}{m_{\Pl}^{2}}  \int_{u_{\text{min}}}^{u_{\text{max}}} \,
\left\{f\bigl(R_{\n}\bigr) +2{R'}^2 f_{RRR}(R) +2\Bigl[R'' +R' \Big(3 \gamma' +(\n-1 ) \dfrac{r'}{r}\Big)  \Bigr] f_{RR}(R)  \right.  
\nonumber \\
& \ \ \ \left. - 2\left( \gamma'' +4{\gamma'}^2 + (\n-1)\frac{\gamma' r'}{r} \right)  f_{R}(R)  
+ \frac{6H^2}{\e^{2\gamma(u)}} f_{R} - f(R) \right\}  \,\e^{4\gamma} r^{\n-1} \, du =0.
 \end{align}
  Part of this expression can be transformed as follows:
\begin{eqnarray} \label{boundary_}
&& \hspace{-10mm}
\mathcal{V}_{\n-1} \frac{m_{\D}^{D-2}}{m_{\Pl}^{2}}\int_{u_{\text{min}}}^{u_{\text{max}}} \,
\left\{ \biggl( 2{R'}^2 f_{RRR} +2\Bigl[R'' +R' \Big(3 \gamma' +(\n-1 ) \dfrac{r'}{r}\Big)  \Bigr] f_{RR} - 2\left( \gamma'' +4{\gamma'}^2  \right. \right.
\nn && \left. \left.
+ (\n-1)\frac{\gamma' r'}{r} \right)  f_{R} \right\} \,\e^{4\gamma} r^{\n-1} \, du
= \mathcal{V}_{\n-1} \frac{m_{\D}^{D-2}}{m_{\Pl}^{2}} 2 \int \Big[ \Big(f_{RR} R' -f_R \gamma' \Big) \,\e^{4\gamma} r^{\n-1}  \Big]' \, du.
\end{eqnarray}
  The remaining part of the expression \eqref{intL_} can be rewritten in a more conventional 
  form (by substituting the expansion \eqref{ser} and the definition \eqref{R4}):
\bear \label{firs}
&&\mathcal{V}_{\n-1} \frac{m_{\D}^{D-2}}{m_{\Pl}^{2}} \int
\left(f(R_\n) +\frac{6H^2}{\e^{2\gamma}} f_{R}(R) - f(R) \right)  \,\e^{4\gamma} r^{\n-1} \, du 
\nn && \cm
= \mathcal{V}_{\n-1} \frac{m_{\D}^{D-2}}{m_{\Pl}^{2}} \int 
\left[f(R_\n) +\frac{6H^2}{\e^{2\gamma(u)}} \left( f_{R}(R_n) +\frac{R_4}{\e^{2\gamma(u)}} f_{RR}(R_n)  \right) - \Big( f(R_n)   \right. 
\nn && \inch  \left.  
+\frac{R_4}{\e^{2\gamma(u)}} f_{R}(R_n) + \frac{R_4^2}{2\e^{4\gamma(u)}} f_{RR}(R_n) \Big) +O\left( \frac{R_4^3}{\e^{6\gamma(u)}} f_{RRR}(R_n) \right) \right]  \,\e^{4\gamma} r^{\n-1} \, du
 \nn && \cm
= \mathcal{V}_{\n-1} \frac{m_{\D}^{D-2}}{m_{\Pl}^{2}} \int
\left[ \frac{(6H^2 -R_4)}{\e^{2\gamma}} f_R(R_n) 
+\frac{(12H^2 -R_4)R_4}{2\e^{4\gamma}} f_{RR}(R_n)\right. 
\nn && \inch  \left.
+O\left( \frac{R_4^3}{\e^{6\gamma(u)}} f_{RRR}(R_n) \right) \right]  \,\e^{4\gamma} r^{\n-1} \, du. 
\ear
  Since $m_{\Pl}^{2}$ is defined by the expression \eqref{mPl} and $R_4=12H^2$, \eq \eqref{firs}}
  can be rewritten as
\beq\label{6H2}
        -6 H^2 + O(H^6). 
\eeq
  Thus we can present $c_{\eff}$ in the form of \eqref{ceff_2} by summing the 
  expressions \eqref{boundary_} and \eqref{6H2}.

\printbibliography[title={References}, heading=bibintoc]

@string{gc="Grav. \& Cosm. "}

@string{jcap="J. Cosmol. Astropart. Phys. "}

@string{jhep="J. High Energ. Phys. "}

@string{prd="Phys. Rev.~{\bf{D}} "}

@string{prl="Phys. Rev. Lett. "}

@article{Bronnikov:2007kw,
    author = "Bronnikov, K. A. and Meierovich, B. E.",
    title = "{Global strings in extra dimensions: A Full map of solutions, matter trapping and the hierarchy problem}",
    eprint = "0708.3439",
    archivePrefix = "arXiv",
    primaryClass = "hep-th",
    doi = "10.1007/s11447-008-2005-0",
    journal = "J. Exp. Theor. Phys.",
    volume = "106",
    pages = "247--264",
    year = "2008"
}

@article{Planck:2018jri,
    author = "Akrami, Y. and others",
    collaboration = "Planck",
    title = "{Planck 2018 results. X. Constraints on inflation}",
    eprint = "1807.06211",
    archivePrefix = "arXiv",
    primaryClass = "astro-ph.CO",
    doi = "10.1051/0004-6361/201833887",
    journal = "Astron. Astrophys.",
    volume = "641",
    pages = "A10",
    year = "2020"
}

@article{Petriakova:2023klf,
    author = "Petriakova, Polina and Popov, Arkady A. and Rubin, Sergey G.",
    title = "{Flexible extra dimensions}",
    eprint = "2303.04785",
    archivePrefix = "arXiv",
    primaryClass = "gr-qc",
    doi = "10.1140/epjc/s10052-023-11542-7",
    journal = "Eur. Phys. J. C",
    volume = "83",
    number = "5",
    pages = "371",
    year = "2023"
}

@article{Marian:2021tje,
    author = "Marian, I. G. and Jentschura, U. D. and Defenu, N. and Trombettoni, A. and Nandori, I.",
    title = "{Vacuum energy and renormalization of the field-independent term}",
    eprint = "2107.06069",
    archivePrefix = "arXiv",
    primaryClass = "hep-th",
    doi = "10.1088/1475-7516/2022/03/062",
    journal = "JCAP",
    volume = "03",
    number = "03",
    pages = "062",
    year = "2022"
}

@article{Arbuzov:2021yai,
    author = "Arbuzov, A. and Latosh, B. and Nikitenko, A.",
    title = "{Effective potential of scalar-tensor gravity with quartic self-interaction of scalar field}",
    eprint = "2109.09797",
    archivePrefix = "arXiv",
    primaryClass = "gr-qc",
    doi = "10.1088/1361-6382/ac4827",
    journal = "Class. Quant. Grav.",
    volume = "39",
    number = "5",
    pages = "055003",
    year = "2022"
}

@article{Liebling:2012fv,
    author = "Liebling, Steven L. and Palenzuela, Carlos",
    title = "{Dynamical boson stars}",
    eprint = "1202.5809",
    archivePrefix = "arXiv",
    primaryClass = "gr-qc",
    doi = "10.1007/s41114-023-00043-4",
    journal = "Living Rev. Rel.",
    volume = "15",
    pages = "6",
    year = "2012"
}

@article{Nikulin:2021bmw,
    author = "Nikulin, Valery V. and Rubin, Sergey G.",
    title = "{Cosmological baryon/lepton asymmetry in terms of Kaluza\textendash{}Klein extra dimensions}",
    eprint = "2109.05469",
    archivePrefix = "arXiv",
    primaryClass = "hep-ph",
    doi = "10.1142/S0218271821400046",
    journal = "Int. J. Mod. Phys. D",
    volume = "30",
    number = "16",
    pages = "2140004",
    year = "2021"
}

@article{RomeroCastellanos:2018inv,
    author = "Romero Castellanos, Ana R. and Sobreira, Flavia and Shapiro, Ilya L. and Starobinsky, Alexei A.",
    title = "{On higher derivative corrections to the $R+R^2$ inflationary model}",
    eprint = "1810.07787",
    archivePrefix = "arXiv",
    primaryClass = "gr-qc",
    doi = "10.1088/1475-7516/2018/12/007",
    journal = "JCAP",
    volume = "12",
    pages = "007",
    year = "2018"
}

@article{Wetterich:2001kra,
    author = "Wetterich, Christof",
    editor = "Horvath, Z. and Palla, L.",
    title = "{Effective average action in statistical physics and quantum field theory}",
    eprint = "hep-ph/0101178",
    archivePrefix = "arXiv",
    reportNumber = "HD-THEP-01-2",
    doi = "10.1142/S0217751X01004591",
    journal = "Int. J. Mod. Phys. A",
    volume = "16",
    pages = "1951--1982",
    year = "2001"
}

@article{Arbuzova:2021etq,
    author = "Arbuzova, Elena and Dolgov, Alexander and Singh, Rajnish",
    title = "{$R^2$-Cosmology and New Windows for Superheavy Dark Matter}",
    doi = "10.3390/sym13050877",
    journal = "Symmetry",
    volume = "13",
    number = "5",
    pages = "877",
    year = "2021"
}

@article{Fabris:2019ecx,
    author = "Fabris, J\'ulio C. and Popov, Arkady A. and Rubin, Sergey G.",
    title = "{Multidimensional gravity with higher derivatives and inflation}",
    eprint = "1911.03695",
    archivePrefix = "arXiv",
    primaryClass = "gr-qc",
    doi = "10.1016/j.physletb.2020.135458",
    journal = "Phys. Lett. B",
    volume = "806",
    pages = "135458",
    year = "2020"
}

@article{Workman:2022ynf,
    author = "Workman, R. L. and Others",
    collaboration = "Particle Data Group",
    title = "{Review of Particle Physics}",
    doi = "10.1093/ptep/ptac097",
    journal = "PTEP",
    volume = "2022",
    pages = "083C01",
    year = "2022"
}

@article{Sui:2017gyi,
    author = "Sui, Tao-Tao and Zhao, Li and Zhang, Yu-Peng and Xie, Qun-Ying",
    title = "{Localization and mass spectra of various matter fields on Weyl thin brane}",
    eprint = "1701.04957",
    archivePrefix = "arXiv",
    primaryClass = "gr-qc",
    doi = "10.1140/epjc/s10052-017-4922-6",
    journal = "Eur. Phys. J. C",
    volume = "77",
    number = "6",
    pages = "411",
    year = "2017"
}

@article{Sorkhi:2018nln,
    author = "Sorkhi, Masoumeh Moazze and Ghalenovi, Zahra",
    title = "{Fermion Localization on the Deformed Brane with the Derivative Coupling Mechanism}",
    doi = "10.5506/APhysPolB.49.123",
    journal = "Acta Phys. Polon. B",
    volume = "49",
    pages = "123--144",
    year = "2018"
}

@article{Arai:2018hao,
    author = "Arai, Masato and Blaschke, Filip and Eto, Minoru and Sakai, Norisuke",
    title = "{Massless bosons on domain walls: Jackiw-Rebbi-like mechanism for bosonic fields}",
    eprint = "1811.08708",
    archivePrefix = "arXiv",
    primaryClass = "hep-th",
    reportNumber = "YGHP-18-09",
    doi = "10.1103/PhysRevD.100.095014",
    journal = "Phys. Rev. D",
    volume = "100",
    number = "9",
    pages = "095014",
    year = "2019"
}

@article{Chumbes:2011zt,
    author = "Chumbes, A. E. R. and Hoff da Silva, J. M. and Hott, M. B.",
    title = "{A model to localize gauge and tensor fields on thick branes}",
    eprint = "1108.3821",
    archivePrefix = "arXiv",
    primaryClass = "hep-th",
    doi = "10.1103/PhysRevD.85.085003",
    journal = "Phys. Rev. D",
    volume = "85",
    pages = "085003",
    year = "2012"
}

@article{Schwinger:1951nm,
    author = "Schwinger, Julian S.",
    editor = "Milton, K. A.",
    title = "{On gauge invariance and vacuum polarization}",
    doi = "10.1103/PhysRev.82.664",
    journal = "Phys. Rev.",
    volume = "82",
    pages = "664--679",
    year = "1951"
}

@article{Rubin:2020pqu,
    author = "Rubin, Sergey G.",
    title = "{How to make the physical parameters small}",
    eprint = "2004.12798",
    archivePrefix = "arXiv",
    primaryClass = "hep-th",
    doi = "10.1155/2020/1048585.",
    journal = "Adv. High Energy Phys.",
    volume = "2020",
    pages = "1048585",
    year = "2020"
}

@article{Guth:1980zm,
    author = "Guth, Alan H.",
    editor = "Fang, Li-Zhi and Ruffini, R.",
    title = "{The Inflationary Universe: A Possible Solution to the Horizon and Flatness Problems}",
    reportNumber = "SLAC-PUB-2576",
    doi = "10.1103/PhysRevD.23.347",
    journal = "Phys. Rev. D",
    volume = "23",
    pages = "347--356",
    year = "1981"
}

@article{Nojiri_2017,
    author = "Nojiri, S. and Odintsov, S. D. and Oikonomou, V. K.",
    title = "{Modified Gravity Theories on a Nutshell: Inflation, Bounce and Late-time Evolution}",
    eprint = "1705.11098",
    archivePrefix = "arXiv",
    primaryClass = "gr-qc",
    reportNumber = "PHYS.REPT.-692-(2017)-1-104, Phys.Rept. 692 (2017) 1-104",
    doi = "10.1016/j.physrep.2017.06.001",
    journal = "Phys. Rept.",
    volume = "692",
    pages = "1--104",
    year = "2017"
}

@ARTICLE{2007PhLB..651..224N,
   author = {{Nojiri}, S. and {Odintsov}, S.~D. and {Tretyakov}, P.~V.},
    title = "{Dark energy from modified F(R)-scalar-Gauss Bonnet gravity}",
  journal = {Physics Letters B},
archivePrefix = "arXiv",
   eprint = {0704.2520},
 primaryClass = "hep-th",
     year = 2007,
    month = jul,
   volume = 651,
    pages = {224-231},
      doi = {10.1016/j.physletb.2007.06.029},
   adsurl = {http://adsabs.harvard.edu/abs/2007PhLB..651..224N}
}

@article{DeFelice:2010aj,
      author         = "De Felice, Antonio and Tsujikawa, Shinji",
      title          = "{f(R) theories}",
      journal        = "Living Rev. Rel.",
      volume         = "13",
      year           = "2010",
      pages          = "3",
      doi            = "10.12942/lrr-2010-3",
      eprint         = "1002.4928",
      archivePrefix  = "arXiv",
      primaryClass   = "gr-qc",
      SLACcitation   = "%%CITATION = ARXIV:1002.4928;%%"
}

@article{Dudas:2005gi,
      author         = "Dudas, E. and Papineau, C. and Rubakov, V. A.",
      title          = "{Flowing to four dimensions}",
      journal        = jhep,
      volume         = "03",
      year           = "2006",
      pages          = "085",
      doi            = "10.1088/1126-6708/2006/03/085",
      eprint         = "hep-th/0512276",
      archivePrefix  = "arXiv",
      primaryClass   = "hep-th",
      reportNumber   = "CERN-PH-TH-2005-267, CPHT-RR-074-1205, LPT-ORSAY-05-86,
                        INR-TH-26-2005",
      SLACcitation   = "%%CITATION = HEP-TH/0512276;%%"
}

@article{Loeb:2006en,
      author         = "Loeb, Abraham",
      title          = "{An Observational Test for the Anthropic Origin of the
      Cosmological Constant}",
      journal        = jcap,
      volume         = "0605",
      year           = "2006",
      pages          = "009",
      doi            = "10.1088/1475-7516/2006/05/009",
      eprint         = "astro-ph/0604242",
      archivePrefix  = "arXiv",
      primaryClass   = "astro-ph",
      SLACcitation   = "%%CITATION = ASTRO-PH/0604242;%%"
}

@article{Ashoorioon:2013eia,
      author         = "Ashoorioon, Amjad and Dimopoulos, Konstantinos and Sheikh-Jabbari, M. M. and Shiu, Gary",
      title          = "{Reconciliation of High Energy Scale Models of Inflation with Planck}",
      journal        = jcap,
      volume         = "1402",
      year           = "2014",
      pages          = "025",
      doi            = "10.1088/1475-7516/2014/02/025",
      eprint         = "1306.4914",
      archivePrefix  = "arXiv",
      primaryClass   = "hep-th",
      SLACcitation   = "%%CITATION = ARXIV:1306.4914;%%"
}

@article{Chaichian:2000az,
      author         = "Chaichian, Masud and Kobakhidze, Archil B.",
      title          = "{Mass hierarchy and localization of gravity in extra
                        time}",
      journal        = "Phys. Lett.",
      volume         = "B488",
      year           = "2000",
      pages          = "117-122",
      doi            = "10.1016/S0370-2693(00)00874-1",
      eprint         = "hep-th/0003269",
      archivePrefix  = "arXiv",
      primaryClass   = "hep-th",
      reportNumber   = "HIP-2000-16-TH",
      SLACcitation   = "%%CITATION = HEP-TH/0003269;%%"
}

@article{Abbott:1984ba,
      author         = "Abbott, Richard B. and Barr, Stephen M. and Ellis,
                        Stephen D.",
      title          = "{Kaluza-Klein Cosmologies and Inflation}",
      journal        = "Phys. Rev.",
      volume         = "D30",
      year           = "1984",
      pages          = "720",
      doi            = "10.1103/PhysRevD.30.720",
      reportNumber   = "DOE-ER-40048-03-P4",
      SLACcitation   = "%%CITATION = PHRVA,D30,720;%%"
}

@article{Krause:2000uj,
      author         = "Krause, Axel",
      title          = "{A Small cosmological constant and back reaction of
                        nonfinetuned parameters}",
      journal        = jhep,
      volume         = "09",
      year           = "2003",
      pages          = "016",
      doi            = "10.1088/1126-6708/2003/09/016",
      eprint         = "hep-th/0007233",
      archivePrefix  = "arXiv",
      primaryClass   = "hep-th",
      reportNumber   = "HU-EP-00-22",
      SLACcitation   = "%%CITATION = HEP-TH/0007233;%%"
}

@article{Rubin:2015pqa,
      author         = "Rubin, Sergey G.",
      title          = "{Scalar field localization on deformed extra space}",
      journal        = "Eur. Phys. J.",
      volume         = "C75",
      year           = "2015",
      number         = "7",
      pages          = "333",
      doi            = "10.1140/epjc/s10052-015-3553-z",
      eprint         = "1503.05011",
      archivePrefix  = "arXiv",
      primaryClass   = "gr-qc",
      SLACcitation   = "%%CITATION = ARXIV:1503.05011;%%"
}

@article{Rubin:2014ffa,
      author         = "Rubin, S. G.",
      title          = "{The role of initial conditions in the universe
      formation}",
      journal        = gc,
      volume         = "21",
      year           = "2015",
      pages          = "143-151",
      doi            = "10.1134/S0202289315020103",
      eprint         = "1403.2062",
      archivePrefix  = "arXiv",
      primaryClass   = "gr-qc",
      SLACcitation   = "%%CITATION = ARXIV:1403.2062;%%"
}

@article{Starobinsky:1980te,
      author         = "Starobinsky, Alexei A.",
      title          = "{A New Type of Isotropic Cosmological Models Without
                        Singularity}",
      journal        = "Phys. Lett.",
      volume         = "B91",
      year           = "1980",
      pages          = "99-102",
      doi            = "10.1016/0370-2693(80)90670-X",
      SLACcitation   = "%%CITATION = PHLTA,B91,99;%%"
}

@article{Gani:2014lka,
      author         = "Gani, Vakhid A. and Dmitriev, Alexander E. and Rubin,
                        Sergei G.",
      title          = "{Deformed compact extra space as dark matter candidate}",
      journal        = "Int. J. Mod. Phys.",
      volume         = "D24",
      year           = "2015",
      pages          = "1545001",
      doi            = "10.1142/S0218271815450017",
      eprint         = "1411.4828",
      archivePrefix  = "arXiv",
      primaryClass   = "gr-qc",
      SLACcitation   = "%%CITATION = ARXIV:1411.4828;%%"
}

@article{Randall:1999vf,
      author         = "Randall, Lisa and Sundrum, Raman",
      title          = "{An Alternative to compactification}",
      journal        = "Phys. Rev. Lett.",
      volume         = "83",
      year           = "1999",
      pages          = "4690-4693",
      doi            = "10.1103/PhysRevLett.83.4690",
      eprint         = "hep-th/9906064",
      archivePrefix  = "arXiv",
      primaryClass   = "hep-th",
      reportNumber   = "MIT-CTP-2874, PUPT-1867, BUHEP-99-13",
      SLACcitation   = "%%CITATION = HEP-TH/9906064;%%"
}

@article{Babic:2001vv,
      author         = "Babic, A. and Guberina, B. and Horvat, R. and Stefancic,
                        H.",
      title          = "{Renormalization group running of the cosmological
                        constant and its implication for the Higgs boson mass in
                        the standard model}",
      journal        = "Phys. Rev.",
      volume         = "D65",
      year           = "2002",
      pages          = "085002",
      doi            = "10.1103/PhysRevD.65.085002",
      eprint         = "hep-ph/0111207",
      archivePrefix  = "arXiv",
      primaryClass   = "hep-ph",
      reportNumber   = "IRB-TH-12-01",
      SLACcitation   = "%%CITATION = HEP-PH/0111207;%%"
}

@article{Brown:2013fba,
      author         = "Brown, Adam R. and Dahlen, Alex and Masoumi, Ali",
      title          = "{Compactifying de Sitter space naturally selects a small
                        cosmological constant}",
      journal        = "Phys. Rev.",
      volume         = "D90",
      year           = "2014",
      number         = "12",
      pages          = "124048",
      doi            = "10.1103/PhysRevD.90.124048",
      eprint         = "1311.2586",
      archivePrefix  = "arXiv",
      primaryClass   = "hep-th",
      SLACcitation   = "%%CITATION = ARXIV:1311.2586;%%"
}

@inproceedings{Burgess:2013ara,
      author         = "Burgess, C. P.",
      title          = "{The Cosmological Constant Problem: Why it's hard to get
                        Dark Energy from Micro-physics}",
      booktitle      = "{Proceedings, 100th Les Houches Summer School:
                        Post-Planck Cosmology: Les Houches, France, July 8 -
                        August 2, 2013}",
      year           = "2015",
      url            = "https://inspirehep.net/record/1254422/files/arXiv:1309.4133.pdf",
      pages          = "149-197",
      doi            = "10.1093/acprof:oso/9780198728856.003.0004",
      eprint         = "1309.4133",
      archivePrefix  = "arXiv",
      primaryClass   = "hep-th",
      SLACcitation   = "%%CITATION = ARXIV:1309.4133;%%"
}

@article{Hertzberg:2015bta,
      author         = "Hertzberg, Mark P. and Masoumi, Ali",
      title          = "{Can Compactifications Solve the Cosmological Constant
                        Problem?}",
      journal        = jcap,
      volume         = "1606",
      year           = "2016",
      number         = "06",
      pages          = "053",
      doi            = "10.1088/1475-7516/2016/06/053",
      eprint         = "1509.05094",
      archivePrefix  = "arXiv",
      primaryClass   = "hep-th",
      reportNumber   = "MIT-CTP-4671",
      SLACcitation   = "%%CITATION = ARXIV:1509.05094;%%"
}

@book{Peskin:1995ev,
      author         = "Peskin, Michael E. and Schroeder, Daniel V.",
      title          = "{An Introduction to quantum field theory}",
      url            = "http://www.slac.stanford.edu/spires/find/books/www?cl=QC174.45%3AP4",
      journal        = "Reading, USA: Addison-Wesley (1995) 842 p",
      year           = "1995",
      %ISBN           = "9780201503975, 0201503972",
      SLACcitation   = "%%CITATION = INSPIRE-407703;%%"
}

@article{Tegmark:2005dy,
      author         = "Tegmark, Max and Aguirre, Anthony and Rees, Martin and
                        Wilczek, Frank",
      title          = "{Dimensionless constants, cosmology and other dark
                        matters}",
      journal        = "Phys. Rev.",
      volume         = "D73",
      year           = "2006",
      pages          = "023505",
      doi            = "10.1103/PhysRevD.73.023505",
      eprint         = "astro-ph/0511774",
      archivePrefix  = "arXiv",
      primaryClass   = "astro-ph",
      SLACcitation   = "%%CITATION = ASTRO-PH/0511774;%%"
}

@article{ArkaniHamed:1998rs,
      author         = "Arkani-Hamed, Nima and Dimopoulos, Savas and Dvali, G. R.",
      title          = "{The Hierarchy problem and new dimensions at a
                        millimeter}",
      journal        = "Phys. Lett.",
      volume         = "B429",
      year           = "1998",
      pages          = "263-272",
      doi            = "10.1016/S0370-2693(98)00466-3",
      eprint         = "hep-ph/9803315",
      archivePrefix  = "arXiv",
      primaryClass   = "hep-ph",
      reportNumber   = "SLAC-PUB-7769, SU-ITP-98-13",
      SLACcitation   = "%%CITATION = HEP-PH/9803315;%%"
}

@ARTICLE{2003PhRvD..68f3516B,
    author = {{Bringmann}, T. and {Eriksson}, M. and {Gustafsson}, M.},
    title = "{Cosmological evolution of homogeneous universal extra dimensions}",
    journal = prd,
    eprint = {arXiv:astro-ph/0303497},
    year = 2003,
    month = sep,
    volume = 68,
    number = 6,
    eid = {063516},
    pages = {063516},
    doi = {10.1103/PhysRevD.68.063516},
    adsurl = {http://adsabs.harvard.edu/abs/2003PhRvD..68f3516B}
}

@inproceedings{Bronnikov:2003rf,
      author         = "Bronnikov, K. A. and Melnikov, V. N.",
      title          = "{Conformal frames and D-dimensional gravity}",
      booktitle      = "{International School of Cosmology and Gravitation: 18th Course: The Gravitational Constant: Generalized Gravitational Theories and Experiments: A NATO Advanced Study Institute Erice, Italy, April 30-May 10, 2003}",
      year           = "2003",
      pages          = "39-64",
      doi            = "10.1007/978-1-4020-2242-5_2",
      eprint         = "gr-qc/0310112",
      archivePrefix  = "arXiv",
      primaryClass   = "gr-qc",
      SLACcitation   = "%%CITATION = GR-QC/0310112;%%"
}

@BOOK{KhlopovRubin,
  author = "Khlopov, M.~Yu. and Rubin, S.~G.",
  title  = "Cosmological Pattern of Microphysics in the Inflationary Universe",
  address    = "P.O. Box 17, 3300 AA Dordrecht, The Netherlands",
  publisher  = "Kluwer Academic Publishers",
  year   = 2004,
  numpages   = 263,
  language   = "english"
}

@BOOK{Lindebook,
    author = {{Linde}, A.~D.},
    title = "{Particle Physics and Inflationary Cosmology}",
    year = 1990,
    numpages = 362,
    publisher = {Harwood Academic Publishers, Switzerland}
}

@ARTICLE{2014JCAP...01..008B,
   author = {{Bamba}, K. and {Makarenko}, A.~N. and {Myagky}, A.~N. and {Nojiri}, S. and
	{Odintsov}, S.~D.},
    title = "{Bounce cosmology from F(R) gravity and F(R) bigravity}",
  journal = jcap,
archivePrefix = "arXiv",
   eprint = {1309.3748},
 primaryClass = "hep-th",
     year = 2014,
    month = jan,
   volume = 1,
      eid = {008},
    pages = {8},
      doi = {10.1088/1475-7516/2014/01/008},
   adsurl = {http://adsabs.harvard.edu/abs/2014JCAP...01..008B}
}

@article{Sokolowski:2007rd,
      author         = "Sokolowski, Leszek M.",
      title  = "{Metric gravity theories and cosmology:II. Stability of a ground state in f(R) theories}",
      journal        = "Class. Quant. Grav.",
      volume         = "24",
      year           = "2007",
      pages          = "3713-3734",
      doi            = "10.1088/0264-9381/24/14/011",
      eprint         = "0707.0942",
      archivePrefix  = "arXiv",
      primaryClass   = "gr-qc",
      SLACcitation   = "%%CITATION = ARXIV:0707.0942;%%"
}

@article{Brandenberger:2006vv,
      author         = "Brandenberger, Robert H. and Nayeri, Ali and Patil,
                        Subodh P. and Vafa, Cumrun",
      title          = "{String gas cosmology and structure formation}",
      journal        = "Int. J. Mod. Phys.",
      volume         = "A22",
      year           = "2007",
      pages          = "3621-3642",
      doi            = "10.1142/S0217751X07037159",
      eprint         = "hep-th/0608121",
      archivePrefix  = "arXiv",
      primaryClass   = "hep-th",
      reportNumber   = "HUTP-06-A0032",
      SLACcitation   = "%%CITATION = HEP-TH/0608121;%%"
}

@ARTICLE{2002PhRvD..65j5022G,
   author = {{Green}, A.~M. and {Mazumdar}, A.},
    title = "{Dynamics of a large extra dimension inspired hybrid inflation model}",
  journal = prd,
   eprint = {hep-ph/0201209},
     year = 2002,
    month = may,
   volume = 65,
   number = 10,
      eid = {105022},
    pages = {105022},
      doi = {10.1103/PhysRevD.65.105022},
   adsurl = {http://adsabs.harvard.edu/abs/2002PhRvD..65j5022G}
}

@ARTICLE{2003PhRvD..68d4010G,
   author = {{G{\"u}nther}, U. and {Moniz}, P. and {Zhuk}, A.},
    title = "{Nonlinear multidimensional cosmological models with form fields: Stabilization of extra dimensions and the cosmological constant problem}",
  journal = {prd},
   eprint = {hep-th/0303023},
     year = 2003,
    month = aug,
   volume = 68,
   number = 4,
      eid = {044010},
    pages = {044010},
      doi = {10.1103/PhysRevD.68.044010},
   adsurl = {http://adsabs.harvard.edu/abs/2003PhRvD..68d4010G}
}

@ARTICLE{2017JCAP...10..001B,
   author = {{Bronnikov}, K.~A. and {Budaev}, R.~I. and {Grobov}, A.~V. and
	{Dmitriev}, A.~E. and {Rubin}, S.~G.},
    title = "{Inhomogeneous compact extra dimensions}",
  journal = jcap,
archivePrefix = "arXiv",
   eprint = {1707.00302},
 primaryClass = "gr-qc",
     year = 2017,
    month = oct,
   volume = 10,
      eid = {001},
    pages = {001},
      doi = {10.1088/1475-7516/2017/10/001},
   adsurl = {http://adsabs.harvard.edu/abs/2017JCAP...10..001B}
}

@article{Bronnikov:2009zza,
    author = "Bronnikov, K.A. and Rubin, S.G. and Svadkovsky, I.V.",
    doi = "10.1134/S0202289309010083",
    journal = gc,
    pages = "32--33",
    title = "{High-order multidimensional gravity and inflation}",
    volume = "15",
    year = "2009"
}

@article{Bronnikov:2009ai,
    author = "Bronnikov, K.A. and Rubin, S.G. and Svadkovsky, I.V.",
    archivePrefix = "arXiv",
    doi = "10.1103/PhysRevD.81.084010",
    eprint = "0912.4862",
    journal = "Phys.\ Rev.\ D",
    pages = "084010",
    primaryClass = "gr-qc",
    title = "{Multidimensional world, inflation and modern acceleration}",
    volume = "81",
    year = "2010"
}

@article{2005PhRvD..71j4018R,
       author = {{Ringeval}, Christophe and {Peter}, Patrick and {Uzan}, Jean-Philippe},
        title = "{Stability of six-dimensional hyperstring braneworlds}",
      journal = prd,
         year = 2005,
        month = may,
       volume = {71},
       number = {10},
          eid = {104018},
        pages = {104018},
          doi = {10.1103/PhysRevD.71.104018},
archivePrefix = {arXiv},
       eprint = {hep-th/0301172},
 primaryClass = {hep-th},
       adsurl = {https://ui.adsabs.harvard.edu/abs/2005PhRvD..71j4018R},
       langid = "english"}

@article{2000PhRvL..85..240G,
       author = {{Gherghetta}, Tony and {Shaposhnikov}, Mikhail},
        title = "{Localizing Gravity on a Stringlike Defect in Six Dimensions}",
      journal = prl,
         year = 2000,
        month = jul,
       volume = {85},
       number = {2},
        pages = {240-243},
          doi = {10.1103/PhysRevLett.85.240},
archivePrefix = {arXiv},
       eprint = {hep-th/0004014},
 primaryClass = {hep-th},
       adsurl = {https://ui.adsabs.harvard.edu/abs/2000PhRvL..85..240G},
       langid = "english"}

@article{2000PhRvL..84.2564G,
       author = {{Gregory}, Ruth},
        title = "{Nonsingular Global String Compactifications}",
      journal = prl,
         year = 2000,
        month = mar,
       volume = {84},
       number = {12},
        pages = {2564-2567},
          doi = {10.1103/PhysRevLett.84.2564},
archivePrefix = {arXiv},
       eprint = {hep-th/9911015},
 primaryClass = {hep-th},
       adsurl = {https://ui.adsabs.harvard.edu/abs/2000PhRvL..84.2564G},
       langid = "english"}

@article{2005PhRvD..71h4002S,
       author = {{Shimono}, Satsuki and {Chiba}, Takeshi},
        title = "{Numerical solutions of inflating higher dimensional global defects}",
      journal = prd,
         year = 2005,
        month = apr,
       volume = {71},
       number = {8},
          eid = {084002},
        pages = {084002},
          doi = {10.1103/PhysRevD.71.084002},
archivePrefix = {arXiv},
       eprint = {gr-qc/0503064},
 primaryClass = {astro-ph},
       adsurl = {https://ui.adsabs.harvard.edu/abs/2005PhRvD..71h4002S},
       langid = "english"}

@article{2000PhRvD..62d4014O,
       author = {{Olasagasti}, Itsaso and {Vilenkin}, Alexander},
        title = "{Gravity of higher-dimensional global defects}",
      journal = prd,
         year = 2000,
        month = aug,
       volume = {62},
       number = {4},
          eid = {044014},
        pages = {044014},
          doi = {10.1103/PhysRevD.62.044014},
archivePrefix = {arXiv},
       eprint = {hep-th/0003300},
 primaryClass = {hep-th},
       adsurl = {https://ui.adsabs.harvard.edu/abs/2000PhRvD..62d4014O},
       langid = "english"}

@article{2003PhRvD..68b5013C,
       author = {{Cho}, Inyong and {Vilenkin}, Alexander},
        title = "{Gravity of superheavy higher-dimensional global defects}",
      journal = prd,
         year = 2003,
        month = jul,
       volume = {68},
       number = {2},
          eid = {025013},
        pages = {025013},
          doi = {10.1103/PhysRevD.68.025013},
archivePrefix = {arXiv},
       eprint = {hep-th/0304219},
 primaryClass = {hep-th},
       adsurl = {https://ui.adsabs.harvard.edu/abs/2003PhRvD..68b5013C},
       langid = "english"}

@article{2018PhRvD..98d3505L,
       author = {{Liu}, Lei-Hua and {Prokopec}, Tomislav and {Starobinsky}, Alexei A.},
        title = "{Inflation in an effective gravitational model and asymptotic safety}",
      journal = prd,
         year = 2018,
        month = aug,
       volume = {98},
       number = {4},
          eid = {043505},
        pages = {043505},
          doi = {10.1103/PhysRevD.98.043505},
archivePrefix = {arXiv},
       eprint = {1806.05407},
 primaryClass = {gr-qc},
       adsurl = {https://ui.adsabs.harvard.edu/abs/2018PhRvD..98d3505L},
       langid = "english"}

@article{Nikulin:2020nub,
    author = "Nikulin, Valery V. and Petriakova, Polina M. and Rubin, Sergey G.",
    title = "{Formation of Conserved Charge at the de Sitter Space}",
    eprint = "2006.01329",
    archivePrefix = "arXiv",
    primaryClass = "gr-qc",
    doi = "10.3390/particles3020027",
    journal = "Particles",
    volume = "3",
    number = "2",
    pages = "355--363",
    year = "2020",
    langid = "english"}

@article{2021arXiv210908373R,
       author = {{Rubin}, Sergey G. and {Fabris}, Julio C.},
        title = "{Distortion of extra dimensions in the inflationary Multiverse}",
      journal = {arXiv e-prints},
         year = 2021,
        month = sep,
          eid = {arXiv:2109.08373},
        pages = {arXiv:2109.08373},
archivePrefix = {arXiv},
       eprint = {2109.08373},
 primaryClass = {gr-qc},
       adsurl = {https://ui.adsabs.harvard.edu/abs/2021arXiv210908373R},
       langid = "english"}

@ARTICLE{2002PhRvD..66d4014G,
       author = {{G{\"u}nther}, U. and {Moniz}, P. and {Zhuk}, A.},
        title = "{Asymptotical AdS space from nonlinear gravitational models with stabilized extra dimensions}",
      journal = prd,
         year = 2002,
        month = aug,
       volume = {66},
       number = {4},
          eid = {044014},
        pages = {044014},
          doi = {10.1103/PhysRevD.66.044014},
archivePrefix = {arXiv},
       eprint = {hep-th/0205148},
 primaryClass = {hep-th},
       adsurl = {https://ui.adsabs.harvard.edu/abs/2002PhRvD..66d4014G},
       langid = "english"}

@ARTICLE{2022JCAP...03..058I,
       author = {{Ivanov}, Vsevolod R. and {Ketov}, Sergei V. and {Pozdeeva}, Ekaterina O. and {Vernov}, Sergey Yu.},
        title = "{Analytic extensions of Starobinsky model of inflation}",
      journal = jcap,
         year = 2022,
        month = mar,
       volume = {2022},
       number = {3},
          eid = {058},
        pages = {058},
          doi = {10.1088/1475-7516/2022/03/058},
archivePrefix = {arXiv},
       eprint = {2111.09058},
 primaryClass = {gr-qc},
       adsurl = {https://ui.adsabs.harvard.edu/abs/2022JCAP...03..058I},
       langid = "english"}

\end{document}